\documentclass{ws-mpla}
\usepackage{lineno}
\begin{document}

\markboth{C. Macolino on behalf of the GERDA collaboration}
{Results on neutrinoless double beta decay from GERDA Phase I}

%%%%%%%%%%%%%%%%%%%%% Publisher's Area please ignore %%%%%%%%%%%%%%
\catchline{}{}{}{}{}
%%%%%%%%%%%%%%%%%%%%%%%%%%%%%%%%%%%%%%%%%%%%%%%%%%%%%%%%%%%%%%%%%%%

%\title{INSTRUCTIONS FOR TYPESETTING MANUSCRIPTS\\
%USING \TeX\ OR \LaTeX\footnote{For the title, try not to use more than 
%three lines. Typeset the title in 10 pt Times Roman, uppercase and 
%boldface.}
%}
\title{Results on neutrinoless double beta decay from GERDA Phase I
%\\
%USING \TeX\ OR \LaTeX\footnote{For the title, try not to use more than 
%three lines. Typeset the title in 10 pt Times Roman, uppercase and 
%boldface.}
}

\author{\footnotesize CARLA MACOLINO ON BEHALF OF THE GERDA COLLABORATION
%\footnote{
%Typeset names in 8 pt Times Roman, uppercase. Use the footnote to 
%indicate the present or permanent address of the author.}
}

\address{INFN, Laboratori Nazionali del Gran Sasso, S.S. 17 BIS km. 18.910\\
67010 Assergi L'Aquila, Italy
\footnote{E-mail address: carla.macolino@lngs.infn.it}}

\maketitle

\pub{Received (Day Month Year)}{Revised (Day Month Year)}

\begin{abstract}
The GERmanium Detector Array, GERDA, is designed to search for
neutrinoless double beta (0$\nu\beta\beta$) decay of $^{76}$Ge
 and it is installed in the Laboratori Nazionali del
 Gran Sasso (LNGS) of INFN, Italy. 
In this review, 
the detection principle and detector setup of GERDA are 
described. Also, the main physics results by GERDA Phase I, 
are discussed.
They include the measurement of the half-life of
2$\nu\beta\beta$
 decay, the background decomposition of the energy spectrum and the
techniques for the discrimination of the background, based on
 the pulse shape of the signal. In the last part of this paper, the
estimation of a limit on the half-life of
 0$\nu\beta\beta$ (T$^{0\nu}_{1/2} >$ 2.1$\cdot$10$^{25}$ yr at 90\% C.L.) 
and the comparison with previous results are discussed.
GERDA data from Phase I strongly disfavour the recent claim 
of 0$\nu\beta\beta$ discovery, based on data from the
Heidelberg-Moscow experiment.

\end{abstract}

\ccode{PACS 14.60.Pq: 23.40.-s; 21.10.Tg}

\section{Introduction and Science Motivation}
The nature and properties of the neutrino have an important impact
 on our knowledge of the Universe.
Recent results about neutrino flavour oscillations have shown evidence
of non-zero neutrino mass and have provided values of the squared
masses of the neutrino mass eigenstates, $\Delta m^2$. 
Neutrinoless double beta decay (0$\nu\beta\beta$) can give additional
information on 
the possible ``Majorana'' nature of the neutrino, i.e. when each neutrino eigenstate $\nu_i$ coincides with its
anti-particle $\bar{\nu_i}$.  In such a case, lepton number would
be no longer conserved and physics
beyond the Standard Model would be required.  Additionally,
neutrinoless double beta decay can give an indirect measurement on the
absolute mass of neutrinos and shed light to the hierarchy of
neutrino masses. This is very important when compared to 
similar bounds from 
%constrain the absolute neutrino mass scale by comparing 
%bounds from data on 
%cambiato
Cosmic Microwave Background (CMB) and Large Scale
Structures (LSS) in the Universe.
\\Two-neutrino double beta decay (2$\nu\beta\beta$) is a
second-order process in the Standard Model in which a nucleus changes
its atomic number $Z$ by two units, with the emission of two beta particles
and two neutrinos, e.g.:
\begin{equation}
(A,Z) \rightarrow (A,Z+2) + 2e^- + 2{\bar{\nu}}_e.
\end{equation}
Such a decay can be observed for some even-even nuclei when ordinary beta decay is
energetically prohibited. This decay is very rare with typical half-lives
ranging from $\sim$10$^{19}$ to $\sim$10$^{21}$ yr.
Neutrinoless double beta decay can be viewed as the ordinary
two-neutrino double beta decay where no neutrino is emitted in the
final state:
\begin{equation}
(A,Z) \rightarrow (A,Z+2) + 2e^-.
\end{equation}
While the 2$\nu\beta\beta$ process is not forbidden by any
conservation law in the Standard Model, 0$\nu\beta\beta$ can only
occur if the neutrino has a non-zero mass and is a Majorana particle. 
Theoretical models predict that 0$\nu\beta\beta$ could be mediated by  a
light Majorana neutrino. Indeed, within the SU$_L$(2)$\times$U(1)
Standard Model of electroweak interactions, 0$\nu\beta\beta$ can be seen as
the exchange of a virtual neutrino between two neutrons in the
nucleus; the Majorana particle emitted by the first neutron contains two
helicity components: 
a dominant negative one and a very small positive one. 
The latter is seen as an antineutrino when absorbed by the second neutron.
This is only possible
if the neutrino
is not in a pure helicity state (therefore it has mass) and it is
identical to its anti-particle. For these reasons, the observation of
neutrinoless double beta decay would definitely demonstrate 
the Majorana nature of the neutrino. In addition, the effective mass 
of the Majorana neutrino can be measured and 
the hierarchy of the mass eigenstates can be investigated. 
\\The effective Majorana neutrino mass is related to the half-life of
the decay by the following relation:
\begin{equation}
\frac{1}{T^{0\nu}_{1/2}(A,Z)} = F^{0\nu} \cdot |{\cal M}^{0\nu}|^2 \cdot
  \Big |\frac{m_{\beta \beta}}{m_e} \Big |^2,
\end{equation}
where $m_e$ is the electron mass, $F^{0\nu}$ is the phase space factor, ${\cal M}^{0\nu}$ is the nuclear
matrix element (NME) and $m_{\beta \beta}$ is the
effective Majorana electron neutrino mass:
\begin{equation}
m_{\beta \beta} \equiv |U_{e1}|^2 m_1 +  |U_{e2}|^2 m_2e^{i\phi_2} +  |U_{e3}|^2 m_3e^{i\phi_3},
 \end{equation}
where $m_i$ are the masses of the neutrino mass
eigenstates, $U_{ei}$ the elements of the
neutrino mixing matrix and $e^{i\phi_2}$ and $e^{i\phi_3}$ the
relative Majorana CP phase factors.
\\The 0$\nu\beta\beta$ decay can be experimentally observed as a
narrow peak in the end-point of the 2$\nu\beta\beta$ decay energy spectrum,
corresponding to the Q-value (Q$_{\beta\beta}$) of the decay. The
number of counts in the peak would allow to quantify
the decay rate of the process or, in case of no signal, to set a lower
limit on it, via the relation:
\begin{equation}
T^{0\nu}_{1/2} = \frac{ln{2} \cdot N_A} {N^{0\nu}} \cdot
\varepsilon \cdot \epsilon \cdot \frac{k}{M_A}
\label{eq:t12n0nu}
\end{equation}
with $N_A$ the Avogradro's number, $\varepsilon$ the total exposure
(detector mass $\times$ live time), $\epsilon$ the detection
efficiency, $k$ the enrichment fraction of the enriched material
($k$ corresponds to the fraction of $^{76}$Ge atoms ($f_{76}$) in GERDA) and
$M_A$ its atomic mass (75.6 g for $^{76}$Ge). $N^{0\nu}$ is the observed signal strength
or the corresponding upper limit. 
\\The GERDA experiment\cite{letterofintent,proposal} searches for neutrinoless double beta
decay of $^{76}$Ge, in which $^{76}$Ge (Z=32)
would decay into $^{76}$Se (Z=34)
 and two electrons. The detectors implemented in the
GERDA setup are semiconductors made from material with an isotope
fraction of $^{76}$Ge enriched to about 86\% ($^{enr}$Ge),
which acts as both the $\beta\beta$ decay source and a 4$\pi$
detector. The detectors are characterized by a very good energy resolution, which
allows a clear distinction of the neutrinoless double beta peak at
$Q_{\beta\beta}$=2039 keV, which is an energy region that is 
nearly
background-free. 
\\Prior to the latest GERDA results, the best limits for 0$\nu\beta\beta$ decay in $^{76}$Ge were
provided by the Heidelberg-Moscow (HdM)\cite{hdmlimit} and
IGEX\cite{igexlimit}
 enriched $^{76}$Ge experiments, that yielded lower
half-life limits of $T_{1/2} > 1.9\cdot10^{25}$ yr and $T_{1/2}                                                                 
> 1.6\cdot10^{25}$ yr
respectively, corresponding to an upper limit on the effective Majorana
mass of $|m_{\beta\beta}| <$ 0.33$\div$1.35 eV; the range in mass
arises from the estimated uncertainty in the nuclear matrix elements. 
A subgroup of the HdM collaboration claimed the observation of
0$\nu\beta\beta$ with a half-life of $T^{0\nu}_{1/2} =                                                                                
$1.19$^{+0.37}_{-0.23}\cdot$10$^{25}$ yr, corresponding to a range for $|m_{\beta\beta}|$ between 0.24
and 0.58 eV, with a central value of 0.44 eV\cite{hdmclaim}.
In a more sophisticated analysis, the authors found a value for
the half-life $T^{0\nu}_{1/2} =                                                                                              
$2.23$^{+0.44}_{-0.31}\cdot$10$^{25}$ yr\cite{hdmclaim2}, though some
inconsistencies associated to this result have been pointed out in Ref.~\refcite{bernhard}.
\\The aim of the Phase I of the GERDA experiment was to verify the previous results
and to reach a much higher sensitivity
than previous experiments. The plan for GERDA Phase II is to reach the
target sensitivity of $T^{0\nu}_{1/2} = 1.4 \cdot 10^{26}$ yr, with an
increased total mass of the 
enriched material and a reduced background level.
\\The outline of the paper is the following: 
in Sect.~\ref{sec:expsetup} it is described the experimental setup
of GERDA at LNGS; the main results concerning 2$\nu\beta\beta$ and 0$\nu\beta\beta$  
decays are discussed in Sects. \ref{sec:2nbb} and \ref{sec:0nbb}. 
The background characterization is described in
Sec.~\ref{sec:gerdabackground} and, finally, pulse-shape
discrimination, used to disentagle the signal from background
events, is  discussed in Sec.~\ref{sec:psd}.

\section{The GERDA experimental setup}
\label{sec:expsetup}
The detection concept of GERDA is implemented by operating bare
$^{enr}$Ge (Ge detectors enriched in $^{76}$Ge) inside a cryostat
containing cryogenic liquid
argon (LAr), surrounded by an additional shield of ultra-pure
water. Liquid argon, indeed, acts both as the coolant medium for the
$^{enr}$Ge detectors and the shield against external gamma radiation\cite{gerdaexpsetup}. 
The Ge detectors are suspended in the cryostat by an array of strings.
\begin{figure}[t!]
\begin{center}
\includegraphics[width=.68\textwidth]{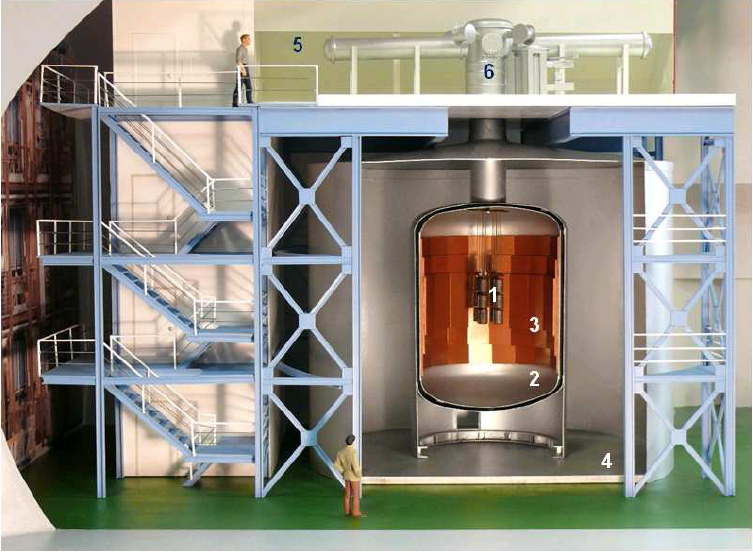}
\caption{An artist's view of the GERDA detector. The array of Ge
  detectors is not to scale. From
  Ref. \protect\refcite{gerdaexpsetup}, with kind permission of The European Physical Journal (EPJ).}
\label{fig:artistview}
\end{center}
\end{figure}
In Fig.~\ref{fig:artistview} an artist's view of the GERDA detector is
shown.
The cryostat is a steel vessel of 4 m diameter with a copper lining,
to reduce gamma radiation from the steel vessel. However,
radon can emanate from the vessel walls and be convected close the
Ge diodes. This can be prevented by separating the central volume 
from the rest of the cryostat by a 3 m high and 750 mm large
cylinder, made of a 30 $\mu$m copper foil (``radon shroud'').
A large tank (8.5 m high and 10 m of diameter) filled with ultra-pure
water surrounds the cryostat and provides a 3 m thick water buffer
around the cryostat. The water buffer is a multi-purpose medium; it is
used to:
(i) moderate and absorb neutrons, (ii) attenuate the flux of external
$\gamma$ radiation, (iii) provide the Cherenkov medium for the
detection of muons and (iv) provide a backup system for the
disposal of the argon gas in case of emergency.
To easily insert the detector strings and the calibration sources into the cryostat,
without increasing the contamination of the cryogenic volume, 
a cleanroom and a lock are located on top of the vessel.
The water tank is instrumented with 66 PMTs, 
to detect Cherenkov light produced 
by muon induced showers in the water buffer. Cherenkov and
scintillation signals, the latter provided by an array of 36 plastic scintillator
panels placed on the roof of the cleanroom, are combined as a muon veto
for the data acquisition according to a logic OR.
\\Data acquisition of GERDA Phase I started on November, 2011 
with 8 p-type $^{enr}$Ge semi-coaxial (HPGe)
detectors, 4 coming from the previous HdM experiment, 1 not enriched
from the GENIUS-Test-Facility\cite{genius} at Gran Sasso and 3 from the
IGEX experiment, with a total mass of about 20.7 kg (17.7 kg
enriched and 3 kg not enriched). 
On July 2012, 
5 Broad Energy GErmanium\footnote{The Broad Energy GErmanium Detectors were manufactured in Olen, Belgium by Canberra.}
(BEGes) diodes, with total mass of about 3.6
kg and foreseen for
the Phase II of the experiment, were also put in place, in order to
test them in a realistic environment. 
The detector array has a structure made of individual strings, each of
them containing up to five independent Ge detectors. 
The energy scale
is determined by calibrating with $^{228}$Th sources on a weekly
basis and was stable during the entire data acquisition period.
Indeed, the differences between the reconstructed peaks of the
$^{228}$Th spectrum
and the ones from the calibration curves are smaller than 0.3
keV. 
In the very first phase of GERDA data taking, a very high background
was observed (18 $\cdot$ 10$^{-2}$ counts/(keV$\cdot$kg$\cdot$
yr)). Also, the line at 1525 keV from $^{42}$K, the progeny of
$^{42}$Ar, had an intensity in the energy spectrum much higher than
expected\cite{ashitkov}. These observations suggested the
hypothesis that charged ions of $^{42}$K drifted in the electric field
produced by the 3 to 4 kV bias of the bare Ge diodes. For this reason
the strings of detectors were enclosed into 60 $\mu$m thick copper
cylinders (``mini-shrouds'').

\section{Measurement of the half-life of 2$\nu\beta\beta$ decay
of $^{76}$Ge with GERDA}
\label{sec:2nbb}
The measurement of the half-life $T^{2\nu}_{\beta\beta}$ of two-neutrino double beta decay
(2$\nu\beta\beta$) of 
$^{76}$Ge is of extreme interest for different
reasons. Firstly, the accurate measurement of
2$\nu\beta\beta$ half-life allows to test the predictions on
${\cal M}^{2\nu}$ 
based on charge exchange experiments\cite{chargeexchange1,chargeexchange2} and, therefore, 
to better understand the nuclear aspects of the decay. Secondly,
as suggested in
Refs.~\refcite{cinquepaper1,cinquepaper2,sei2nu,sei2nu2,sei2nu3,sei2nu4},
the nuclear matrix elements
${\cal M}^{0\nu}$ and ${\cal M}^{2\nu}$, for 0$\nu\beta\beta$ and  2$\nu\beta\beta$ decays respectively, are
related; therefore, the estimation of $T^{2\nu}_{\beta\beta}$ can also constrain the
value of  ${\cal M}^{0\nu}$, that is subject to theoretical uncertainties.
The considered GERDA data set consists of 8796 events, taken between November 2011 and March
2012, from the six enriched semi-coaxial detectors, with a total
collected exposure of 5.04 kg$\cdot$yr. 
The very low energy part of the spectrum is dominated
by the $\beta^{-}$ decay of $^{39}$Ar, produced by cosmogenic
activation of natural argon in the atmosphere, with a Q-value of 565
keV.
At higher energies the spectrum is completely dominated by the
2$\nu\beta\beta$ decay. For this reason, the search for 2$\nu\beta\beta$ decay
the analysis was performed for events with energy in the range between
600 and 1800 keV. According to Monte Carlo simulations, 
 the probability for the 2$\nu\beta\beta$ process to
release energy above 1800 keV in the GERDA detectors is 0.02\%. 
\begin{figure}[t!]
\begin{center}
\includegraphics[width=.58\textwidth]{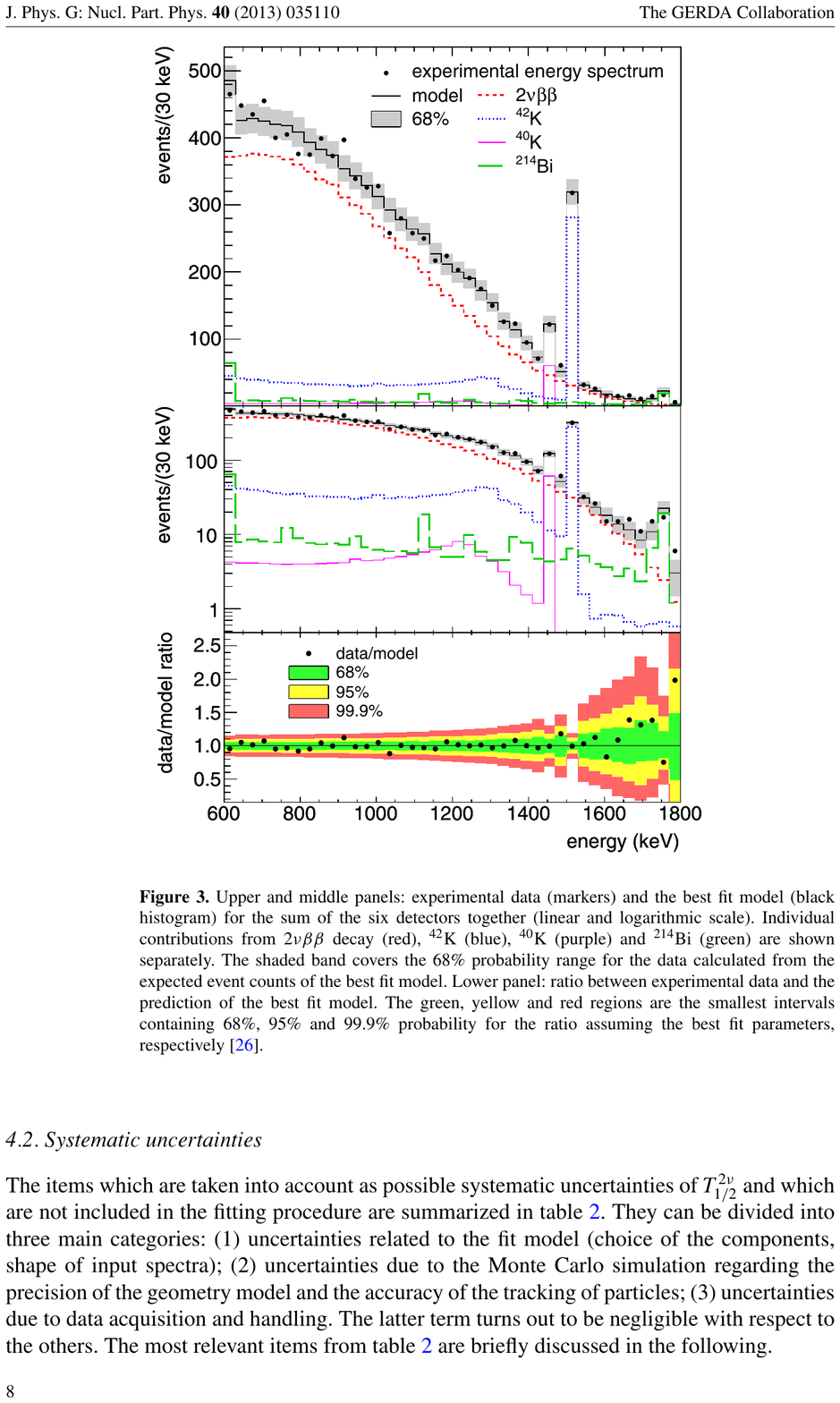}
\caption{Upper and middle panels: Experimental data (markers) and the
  best fit model (black histogram) in linear and logarithmic scale
  (Data refer to the sum of the six detectors). Individual
  contributions from the 2$\nu\beta\beta$ decay (red), $^{42}$K (blue),
$^{40}$K (purple) and $^{214}$Bi (green) are also shown. The shaded band indicates
the 68\% probability range for the data calculated from the expected event counts
of the best fit model. Lower panel: ratio between experimental data and the
prediction of the best fit model as a function of the energy. Plot
from Ref.~\protect\refcite{paper2nu}. \copyright IOP Publishing. Reproduced by permission of IOP Publishing. All rights reserved.}
\label{fig:2nubetabetaresult}
\end{center}
\end{figure}
The analysis of the energy spectra from the six semi-coaxial diodes
was based on a maximum likelihood approach\cite{likelihood}, fitting each
 energy spectrum with a global model which
 considers the 2$\nu\beta\beta$ decay of $^{76}$Ge and three
independent background contributions: $^{42}$K, $^{214}$Bi and
$^{40}$K. $^{42}$K is a progeny of $^{42}$Ar, while $^{214}$Bi
(originated from the $^{238}$U decay series) and $^{40}$K are gamma emitters from the
environmental radioactivity; their gamma lines are observed in the experimental spectrum at 1525
keV, 1764 keV and 1460 keV, respectively. Other background
contributions are not included in the fit, since their gamma lines are
not clearly distinguished in the spectrum, either 
because of low statistical significance or because they are not observed in all
the spectra.
The energy spectra for the
model components (signal plus backgrounds) were obtained from Monte
Carlo simulations, for each 
detector, by using the M{\scriptsize A}G{\scriptsize E}
framework\cite{mage} 
based on G{\scriptsize EANT}4\cite{geant4,geant42}. The
energy distribution of the two electrons was simulated according to the
model\cite{cite3brugnera} implemented in the DECAY0 code\cite{decay0brugnera}.
While the activity of $^{42}$K was assumed to be uniformly distributed in the
LAr volume, $^{40}$K and $^{214}$Bi emitters are assumed as
``close sources'', i.e. located in the detector assembly.
The ratio of the intensities of the $^{214}$Bi $\gamma$ lines observed
in the energy spectrum is consistent with this hypothesis. 
The fit parameters were taken to be the half-life of the 2$\nu\beta\beta$ decay, common to
the six spectra, and the intensities of the background components 
(considered as independent for each detector). Additionally, the active masses
and the $^{76}$Ge abundance of each detector were considered as nuisance parameters
and integrated at the end of the analysis.
The spectral
fit was performed using the Bayesian Analysis Toolkit\cite{bat}. 
The prior probability density function (PDF) for
$T^{2\nu}_{1/2}$ was considered as a flat distribution between 0 and 10$^{22}$ yr, while the prior
PDFs for the active mass fraction and the $^{76}$Ge isotopic abundance
of each detector were modelled according to a
Gaussian distribution.
\\In Fig.~\ref{fig:2nubetabetaresult} the best fit model is shown, together with
experimental data for the sum of the six detectors and the individual
components obtained from the fit. The best
fit model gives an expectation of 8797.0 events divided as follows: 7030.1
(79.9\%) from the 2$\nu\beta\beta$ decay of $^{76}$Ge,
1244.6 (14.1\%) from $^{42}$K, 335.5 (3.8\%) from $^{214}$Bi and 186.8
(2.1\%) from $^{40}$K. The average signal to background ratio is
4:1. The model reproduces very well the experimental data,
with a $p$-value of the fit equal to $p$=0.77.
The best estimate of the derived half-life of 2$\nu\beta\beta$ is
\begin{equation}
 T^{2\nu}_{1/2} =                                                                   
(1.84^{+0.09}_{-0.08 \mbox{ \scriptsize fit}}{\mbox{ }}^{+0.11}_{-0.06\mbox{ \scriptsize syst}})\times10^{21}
\mbox{yr} = (1.84^{+0.14}_{-0.10})\times10^{21}
\mbox{yr},
\end{equation}
where the fit and the systematical uncertainties are summed in quadrature.
\\The fit's error on $T^{2\nu}_{1/2}$ includes both the statistical and
the error associated to the marginalization of the nuisance parameters.
The systematic error on $T^{2\nu}_{1/2}$ 
includes uncertainties related to the background model (position and
distribution of the sources), 
uncertainties due
to Monte Carlo simulation details and errors related to data
acquisition and data handling. The combination in quadrature of all
these contributions gives a systematic uncertainty of
${}^{+6.2}_{-3.3}$\%, which corresponds to $^{+0.11}_{-0.06} \times                                                                   
$10$^{21}$ yr.
\begin{figure}[t!]
\begin{center}
\includegraphics[width=.6\textwidth]{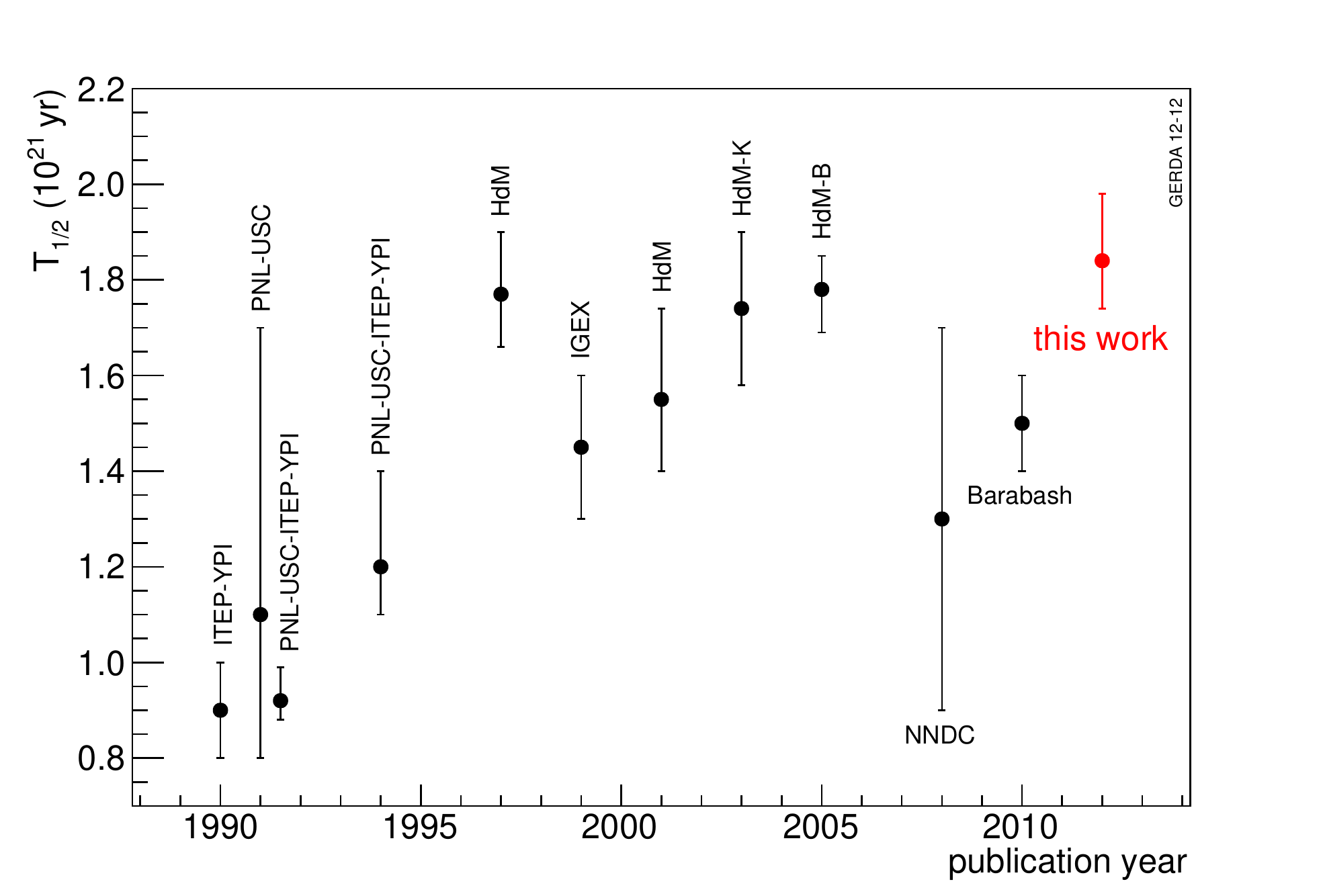}
\caption{T$^{2\nu}_{1/2}$ experimental values for $^{76}$Ge from
  previous experiments compared to the GERDA result. Plot from
  Ref.~\protect\refcite{paper2nu}. \copyright IOP Publishing. Reproduced by permission of IOP Publishing. All rights reserved.}
\label{fig:2vbbresults}
\end{center}
\end{figure}
The comparison between the half-life estimated by GERDA and those from previous
measurements for $^{76}$Ge is shown in
Fig.~\ref{fig:2vbbresults}. 
Almost all the estimated values for T$^{2\nu}_{1/2}$ tend to be larger with the
publication year, probably because 
of an increasing signal-to-background ratio, which makes the effect
of background modelling and subtraction less important. 
with the exstimate given in 
in Ref.~\refcite{ref2result}  (labelled as ``Barabash''); indeed, if T$^{2\nu}_{1/2}$ were as
short as 1.5$\cdot$10$^{21}$ yr, almost all counts detected in the range
600-1800 keV should be due to 2$\nu\beta\beta$ (expected: 8782.7,
observed: 8976), with nearly no
possible contribution from background.
On the other hand, the GERDA estimation is in better agreement with the two most recent results 
(``HdM-K'' and ``HdM-B'')
reported in Refs.~\refcite{result2vbb1} and ~\refcite{result2vbb2}, 
based on the re-analysis of HdM data. 
\\The experimental nuclear matrix 
element for the 2$\nu\beta\beta$
decay of $^{76}$Ge was derived from the measured the half-life, using
the phase space factors from the recently improved electron wave
functions\cite{wavefunctions}:
${\cal M}^{2\nu}$ =
0.133$^{+0.004}_{-0.005}$ MeV$^{-1}$. In
Ref.~\refcite{cinquepaper1} the matrix elements ${\cal M}^{2\nu}$ and
${\cal M}^{0\nu}$ for $^{76}$Ge were related in the QRPA approximation. Notice that, the
half-life measured by GERDA corresponds to a value for ${\cal M}^{2\nu}$ that is  11\% smaller than
the one quoted in ref.~\refcite{cinquepaper1}.
Using that value for  ${\cal M}^{2\nu}$, and the relation between
${\cal M}^{2\nu}$ and ${\cal M}^{0\nu}$ showed in Ref.~\refcite{cinquepaper2}, the predicted half-life for 0$\nu\beta\beta$
 is about 15\% larger but still  well within the uncertainty of the model calculation.  
The nuclear matrix element for 2$\nu\beta\beta$ decay of $^{76}$Ge was
also estimated from charge exchange reactions\cite{chargeexchange1,chargeexchange2}
 (d,$^2$He) and ($^3$He,t). Their value is larger, but still
 consistent, than the one derived by the GERDA
 measurement.

\section{The background of GERDA Phase I}
\label{sec:gerdabackground}
\begin{figure}[!t]
\begin{center}
\includegraphics[width=0.85\textwidth]{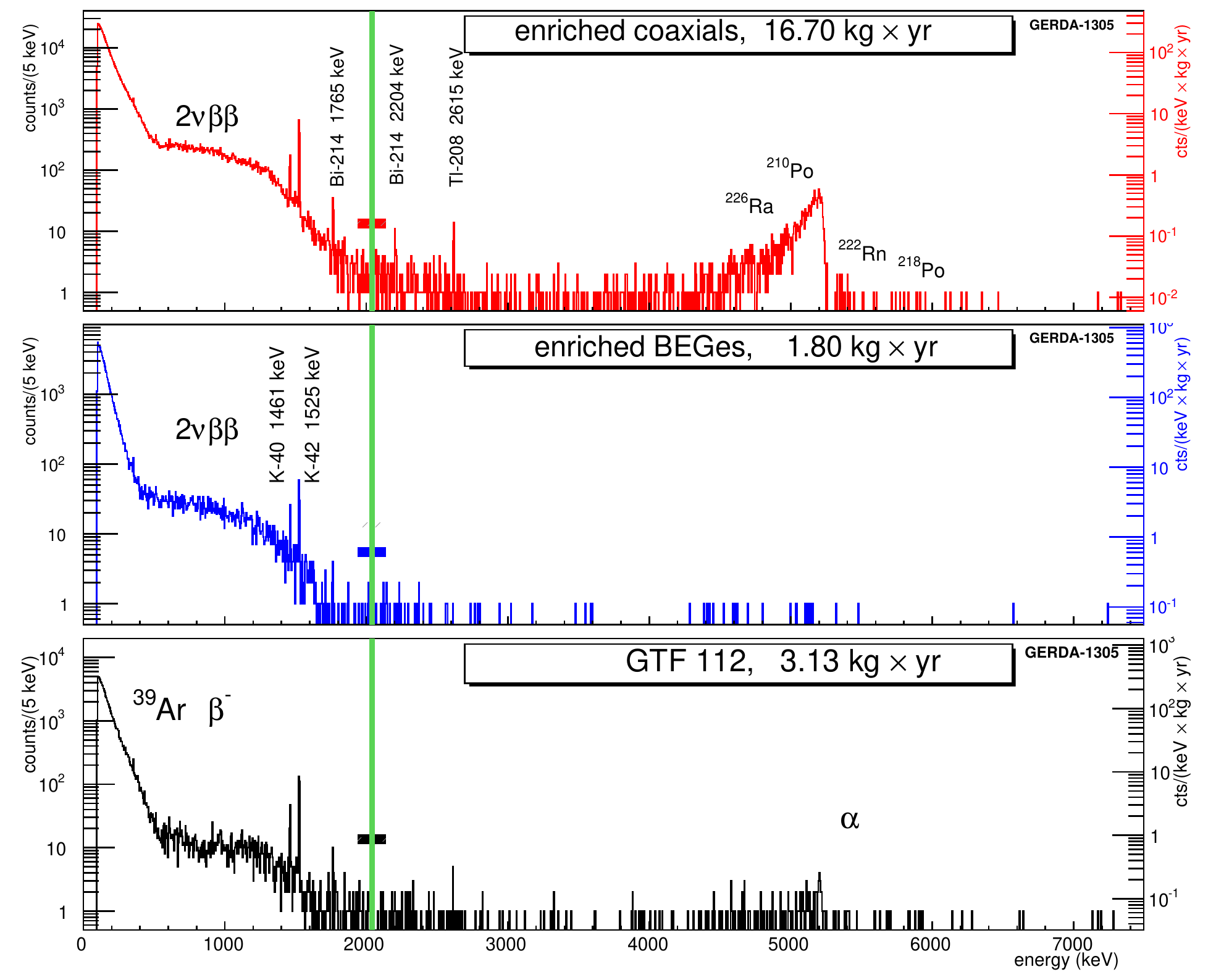}
\caption{Spectra from enriched semi-coaxial (top), enriched BEGe (middle)
  and non-enriched (bottom) detectors of GERDA Phase I. The green line 
indicates the Q$_{\beta\beta} \pm$ 20 keV region of blinded data. The 
bar on the right side of the y-axis indicates the corresponding 
background index. Plots from Ref. \protect\refcite{gerdabackground}, with kind permission of The European Physical Journal (EPJ).}
\label{fig:spectra}
\end{center}
\end{figure}

In order to extract a possible signal at Q$_{\beta\beta}$ or, in case
of no event, to determine a lower limit on the half-life of
the 0$\nu\beta\beta$ process, it is important to understand the
different contributions observed in the energy spectrum. The
identification of these contributions is also crucial to
derive a strategy for background suppression.
The background decomposition of the GERDA energy spectrum was done 
considering data for a total exposure of 16.70 kg$\cdot$yr. The data
were divided according to the different Background Index (BI) in the
region around Q$_{\beta\beta}$, defined
as the number of counts per keV$\cdot$kg$\cdot$yr; in particular, \textit{SILVER-coax}
are data from semi-coaxial detectors collected after the deployment of BEGe detectors in LAr
(1.30 kg$\cdot$yr exposure), \textit{GOLD-coax} are data from semi-coaxial detectors
except \textit{SILVER-coax} data (15.40 kg$\cdot$yr exposure) and \textit{BEGe}
are data from the BEGe detectors \cite{gerdabackground} (1.80 kg$\cdot$yr exposure).
\\The energy spectra from the enriched semi-coaxial detectors, from the BEGes and
from the detector with natural germanium are shown in Fig.~\ref{fig:spectra}. 
In the low energy part, up to 565 keV, the energy spectra are
dominated by the $\beta$-decay of cosmogenic $^{39}$Ar. 
Some differences in the shape of the low energy spectrum is expected
between the semi-coaxial and BEGe detectors because of the slight
difference of geometry and of the n$^+$ dead layer thicknesses.
In the region between 600 and 1500 keV, the enriched detector spectra are dominated
by the continuous spectrum of the 2$\nu\beta\beta$ decay\cite{paper2nu}. 
\\All spectra show $\gamma$ lines from the decay of $^{40}$K and
$^{42}$K, while enriched detectors spectra show lines also from
$^{60}$Co, $^{208}$Tl, $^{214}$Bi, $^{214}$Pb and $^{228}$Ac.
Just a single line from $^{214}$Bi appears clearly in the 
spectrum between 2000 and 2600 keV (at 2204 keV with 17.3 counts). 
Additional $\gamma$
lines from $^{214}$Bi are not expected in this range
 due to the much lower branching ratios of the transitions (the
 strongest line at 2448 keV would give 5.5 counts).
Different peak-like structures appear in the high energy part of the
spectra;  in particular, the  important peak-like structure at 5.3 MeV for the
enriched detectors  can be attributed to the $\alpha$ decay of
$^{210}$Po on the p$^+$ surface of the detectors. 
Additional other
peak-like structures, due the $\alpha$ decays on the p$^+$ surfaces,
appear at 4.7 MeV (for $^{226}$Ra), 5.4 MeV (for $^{222}$Rn) and 5.9
MeV (for $^{218}$Po).
\\Some of the background components identified in the spectra can be
 traced back, from the the material screening, nearby the detector and
 from the electronics. Indeed, $^{228}$Ac and
$^{228}$Th are expected in the front end electronics and in the detector suspension
system, $^{224}$Ra daughters $^{214}$Bi and $^{214}$Pb are expected close to the
detectors, $^{40}$K is expected from the detector assembly and $^{42}$K
from the $\beta$-decay of $^{42}$Ar. 
\begin{figure}[bh!]
\begin{center}
\includegraphics[width=.49\textwidth]{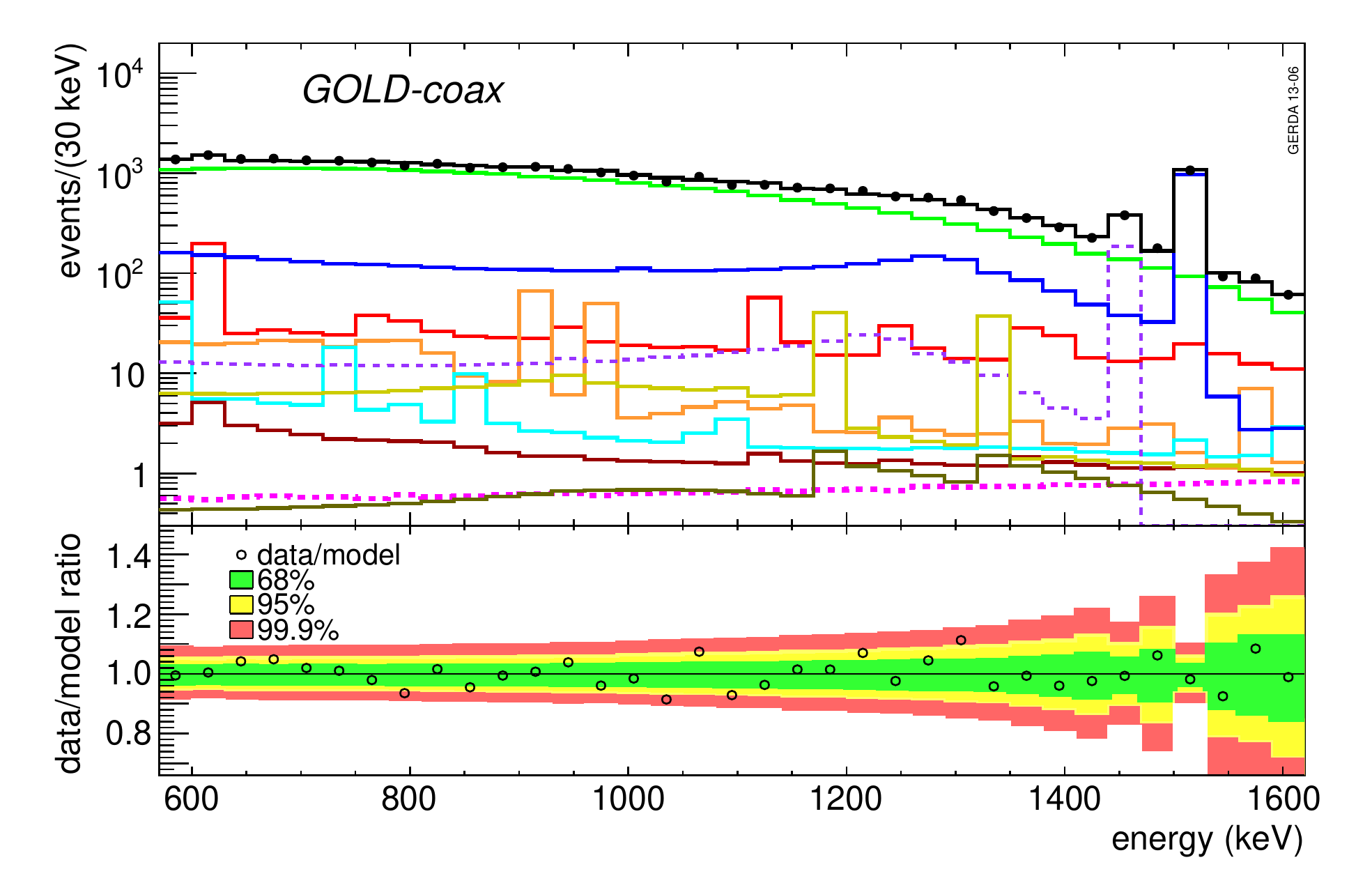}
\includegraphics[width=.49\textwidth]{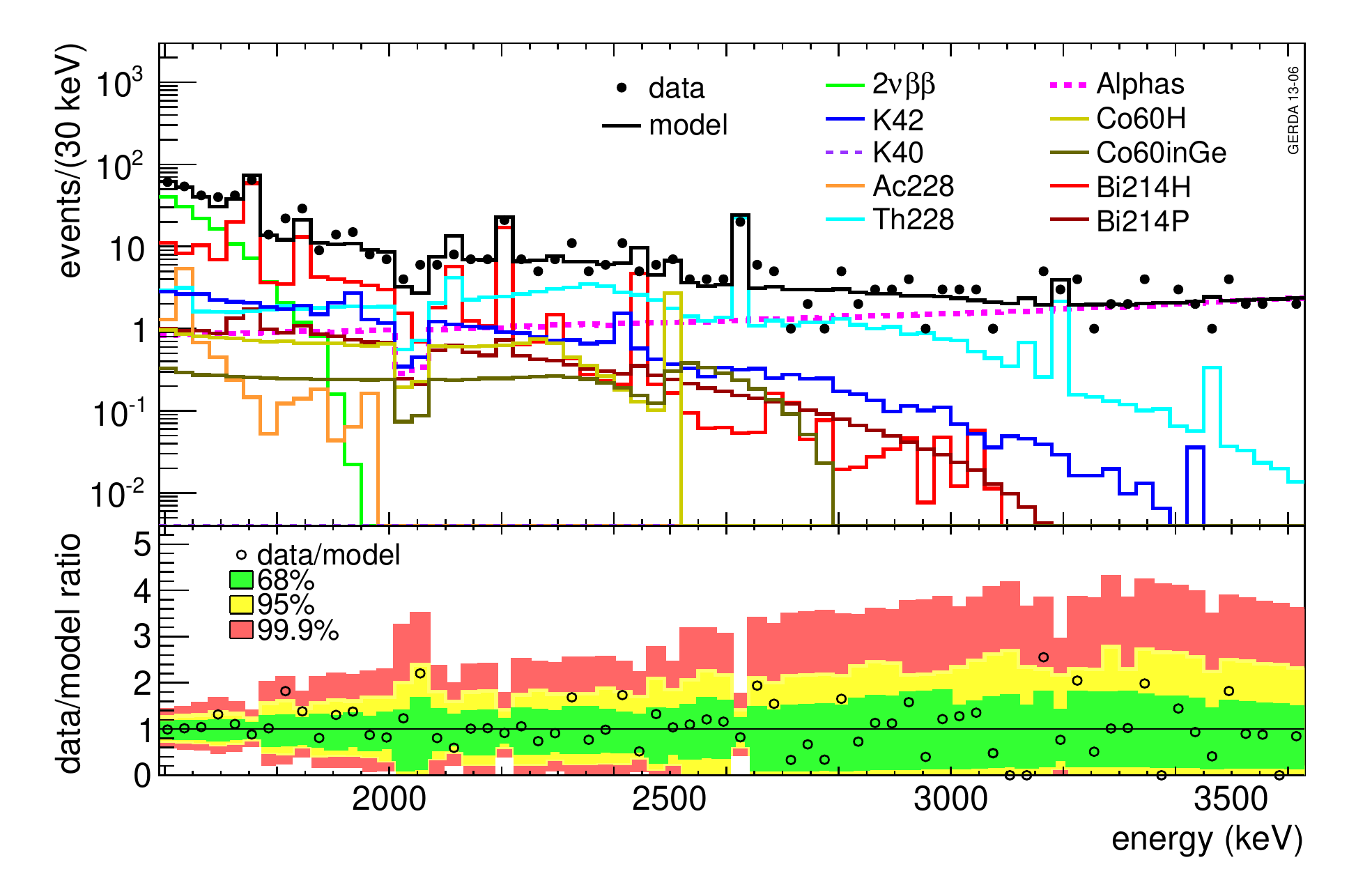}
\\\includegraphics[width=.49\textwidth]{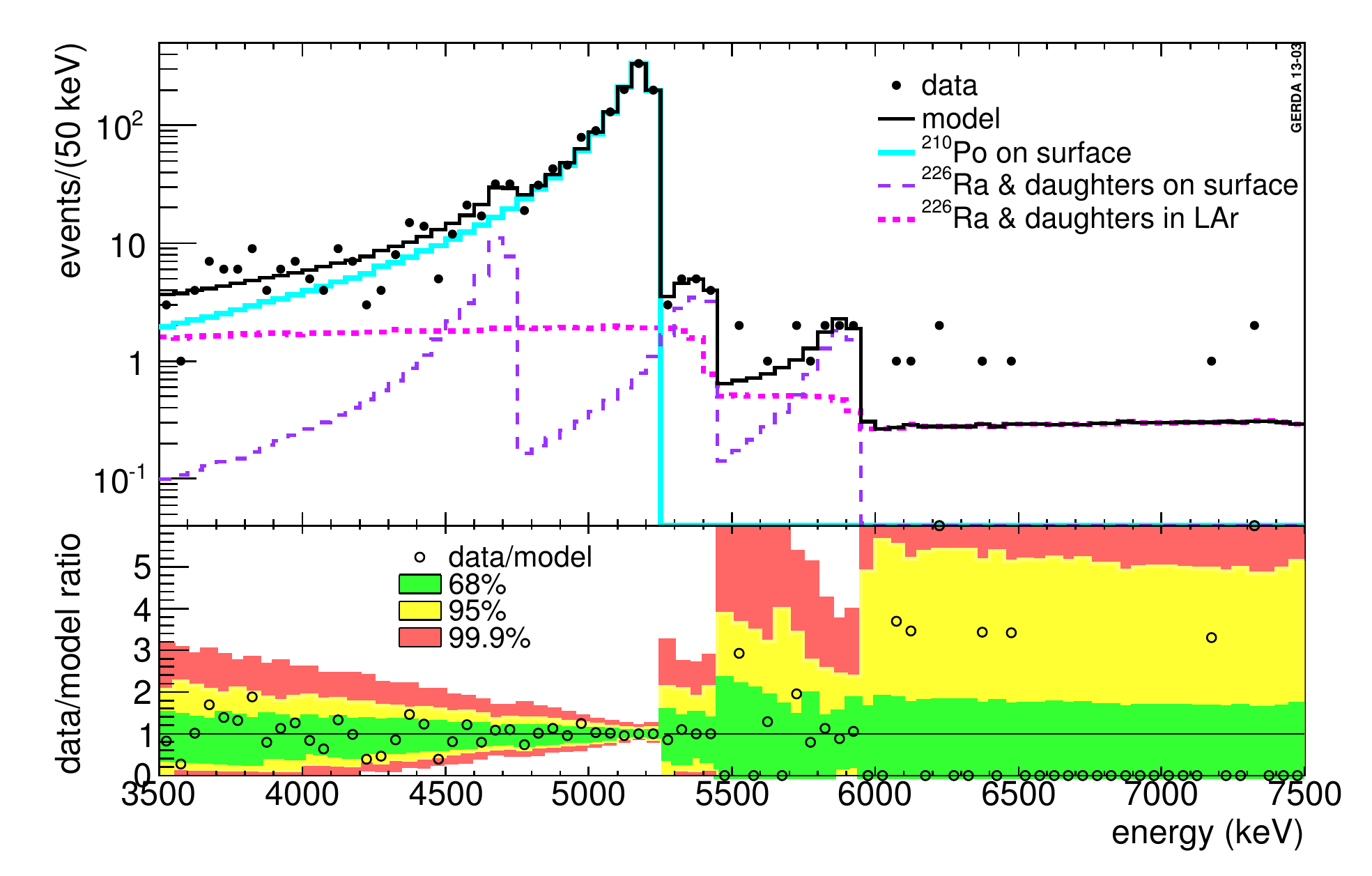}
\caption{Background decomposition for data from \textit{GOLD-coax}
  data set, according to the best fit minimum model. The lower panel
  shows the ratio between experimental data and the prediction from
  the best fit model, togheter with the 68\% (green), 95\% (yellow) and 99.9\% (red)
probability bands. Plots from Ref. \protect\refcite{gerdabackground}, with kind permission of The European Physical Journal (EPJ).}
\label{fig:bckspectra}
\end{center}
\end{figure}
The latter is homogeneously
distributed inside LAr, though $^{42}$K ions can drift in the electric
fields surrounding the detectors.
Neutron and muon fluxes are expected to be of
the order of 10$^{-5}$ cts/(keV$\cdot$kg$\cdot$yr) and 
10$^{-4}$ cts/(keV$\cdot$kg$\cdot$yr) respectively\cite{neutron,muon}
and their contribution can be neglected. Finally, 
isotopes like $^{76}$Ge (via neutron capture), $^{206}$Pb (by
inelastic neutron scattering) and $^{56}$Co (via decay) could also
cause $\gamma$ show up as peaks in the vicinity of Q$_{\beta\beta}$,
however these components either have very
short half-lives or simply are not
associated to other peaks that should be 
observed in the GERDA spectrum.
\\All background components previously discussed were simulated in the
the M{\scriptsize A}G{\scriptsize E} framework\cite{mage} based on
G{\scriptsize EANT}4\cite{geant4,geant42} implementing also 
the GERDA Phase I detectors arrangement  in four  strings. 
The contamination contributions were simulated into the different hardware
components of the detector setup: inside the germanium, on the p$^+$
and n$^+$ surfaces of the detectors, in the liquid argon close to the
p$^+$ surface, homogeneously distributed in the LAr, in the detector
assembly, in the mini-shroud, in the radon shroud and in the heat-exchanger.
The simulated energy spectra were smeared with a Gaussian distribution
with Full Width Half Maximum (FWHM) equal to the detector resolution.
\\Two global models were obtained through a Bayesian fit of the
simulated energy spectrum 
to the measured one: a ``minimum model'' fit, were
 only a minimum amount of background
components were considered, and a ``maximum model'' fit containing all
the possible contributions. In the ``minimum model'' only
background sources located close to the detectors (up to
2cm), were considered. In the ``maximum
model'' further medium and large distance background
components, assumed to be present in different hardware components of
the experiment, were added to the model. 
Once fitted the models to the data, the result was used to derive 
the activities of the different background contributions. It turns out
that data are well described by both models and that there is no unique determination of the
count rates of the different background components. 
However, the largest fraction of background comes
from close sources even when medium and large distance sources are
added, especially on the p+ and n+ surfaces. 
The best fit for the ``minimum model''  and the energy spectrum
of the \textit{GOLD-coax} data set are shown in
Fig.~\ref{fig:bckspectra} for different energies in the range between 570 keV and
3750 keV.  In the energy region between 570 keV and 1500 keV, the
spectrum is well described by the same background components
considered for the 2$\nu\beta\beta$ analysis, i.e. $^{42}$K, $^{40}$K
and $^{214}$Bi.
In the range between 3.5 MeV and 7.5 MeV, the background is
expected to come mainly from $\alpha$ emitting isotopes in the $^{226}$Ra decay
chain, which can be broken at $^{210}$Pb (with half-life of 22.3 yr)
and at $^{210}$Po (with half-life of 138.4 days). The time
distribution of the events confirms the presence of $^{210}$Po, since
data are well fit very by  a decreasing  exponential  plus a
constant  distribution and the Bayesian fit with a non-informative
prior for the half-life gives T$_{1/2}$=130.4$\pm$22.4
days, which is in very good agreement with the half-life of $^{210}$Po.
\begin{figure}[!t]
\begin{center}
\includegraphics[width=.75\textwidth]{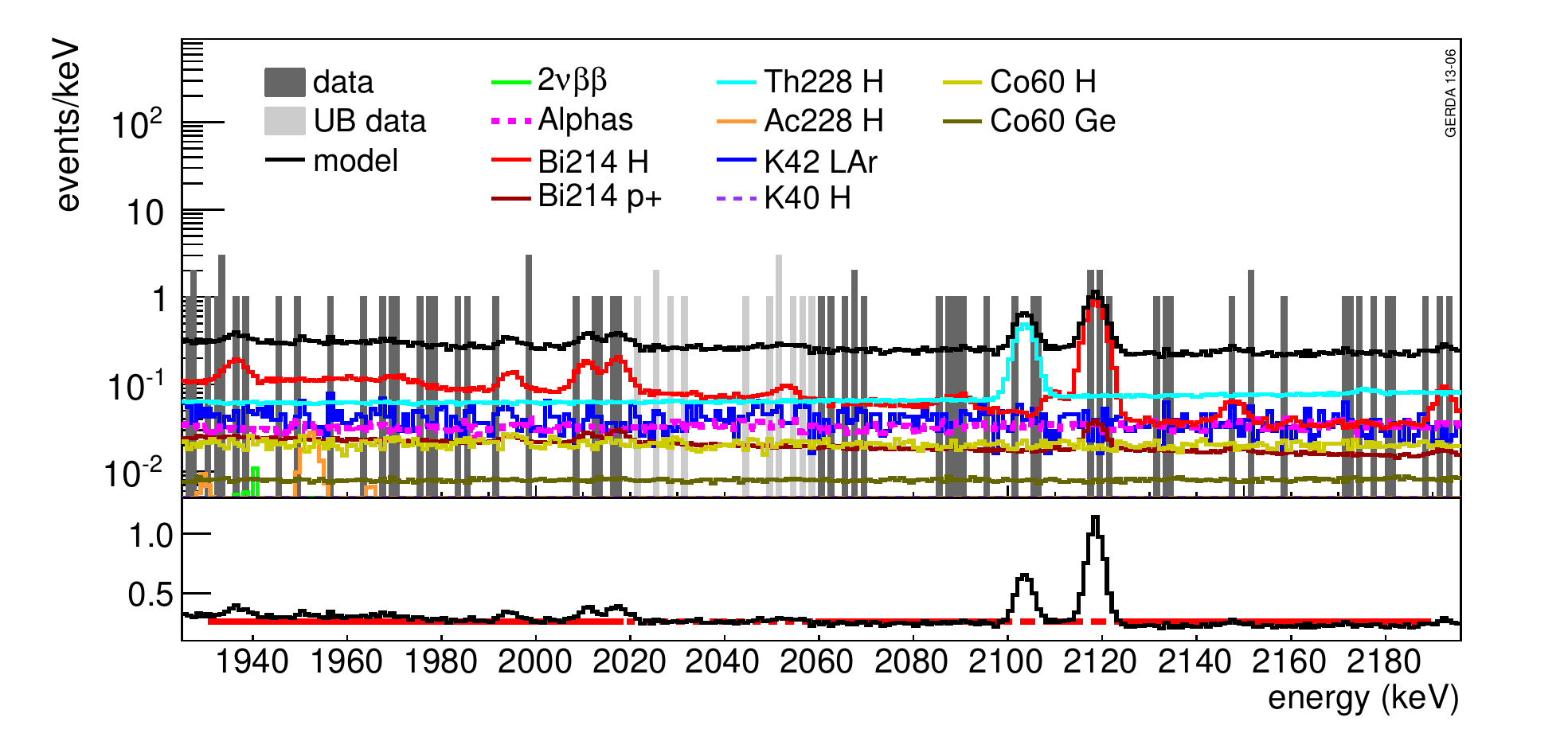}
\\\includegraphics[width=.75\textwidth]{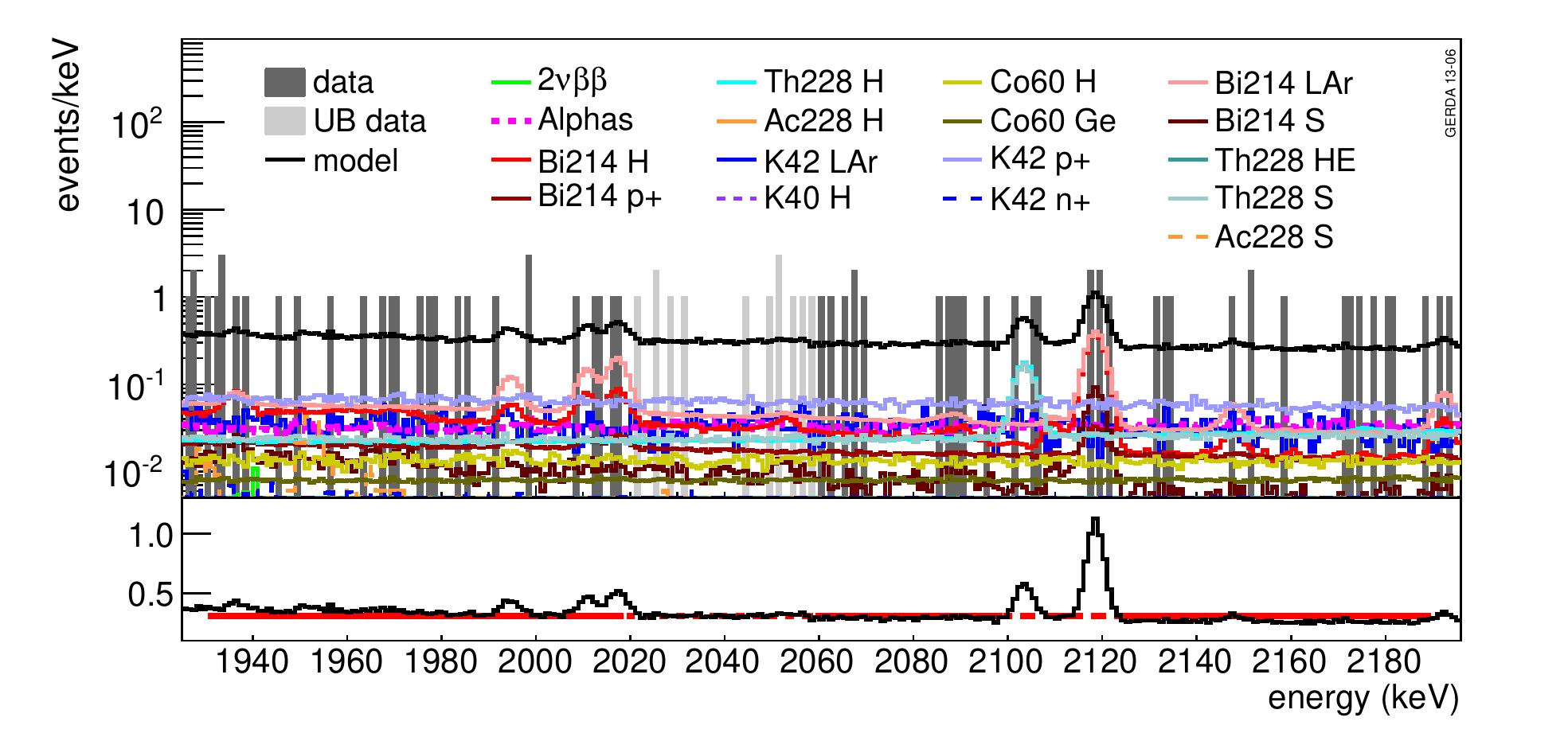}
\caption{Energy spectrum for experimental data together with the
  minimum (top) and maximum (bottom) models around Q$_{\beta\beta}$
  for the \textit{GOLD-coax} data set. The upper panels show the individual
  contributions and the lower panels show the model fitted with a
  constant. Plots from Ref. \protect\refcite{gerdabackground}, with kind permission of The European Physical Journal (EPJ).}
\label{fig:minimummaximum}
\end{center}
\end{figure}
The spectrum is described well
by $^{210}$Po on the surface of the detectors and $^{226}$Ra and its
daughter nuclei on the p$^+$ surface and in liquid argon.
The best fits in the region around the
Q$_{\beta\beta}$ value are shown in figure \ref{fig:minimummaximum},
for both the ``minimum'' and the ``maximum'' model. The predicted background around
Q$_{\beta\beta}$ is flat, with no contribution from $\gamma$ peaks. The
resulting BI for the ``minimum model'' is 1.85$^{+0.08}_{-0.09} \times$
10$^{-2}$ cts/(keV$\cdot$kg$\cdot$yr), while for the ``maximum model'' 
is 2.19$^{+0.19}_{-0.12} \times$ 10$^{-2}$ cts/(keV$\cdot$kg$\cdot$yr). 
The Background Index for \textit{GOLD-coax} data has been evaluated in
the energy window between 1930 and 2190 keV, with the exclusion of the
central 40 keV window around Q$_{\beta\beta}$ and of the $\pm$ 5 keV
regions around the position of $\gamma$ lines, expected from the
background model (single escape peak from $^{208}$Tl at 2104 keV and
$\gamma$ line at 2119 keV from $^{214}$Bi). 
The measured value (BI$=$1.75$^{+0.26}_{-0.24} \times$10$^{-2}$
cts/(keV$\cdot$kg$\cdot$yr)) is in good 
agreement with the values predicted by the two models.
In both cases, the most significant
contributions to the background in the Q$_{\beta\beta}$ region,
 come from $^{214}$Bi and
$^{228}$Th in the detector assembly, from $^{42}$K, homogeneously
distributed in LAr, and from $\alpha$ emitters. 
\\Concerning the background model for  \textit{BEGe} data, only a
qualitatively analysis was possible, since the exposure collected with
these detectors was small. 
Considering the ``minimum model'' fit, with the addition of $^{68}$Ge decays in germanium and
$^{42}$K decays on the n$^+$ surface, the dominant background source
around Q$_{\beta\beta}$ 
is given by $^{42}$K on the n$^+$surface. This contribution is expected to be more important in BEGe
detectors than the semi-coaxial ones because of a thinner dead layer.
\\In order to keep analysis cuts and procedures not biased, GERDA Phase I data 
were blinded (not processed) in a 40 keV energy window between 2019 keV and
2049 keV, up to the collection of 20
kg$\cdot$ yr exposure. Eventually, data were partially unblinded with
a still-blinded window of 10 keV for the \textit{GOLD-coax}
and \textit{SILVER-coax} subsets, 
and 8 keV window for the \textit{BEGe} data set. The models show good agreement
with the observed spectrum around Q$_{\beta\beta}$.
In the unblinded 30 keV window for \textit{GOLD-coax} data, 8.6 events
were predicted by the ``minimum model'' fit and 10.3 by the ``maximum'' one,
while 13 events were observed.

\section{Pulse shape discrimination of GERDA Phase I data}
\label{sec:psd}
\begin{figure}[!t]
\begin{center}
\includegraphics[width=.67\textwidth]{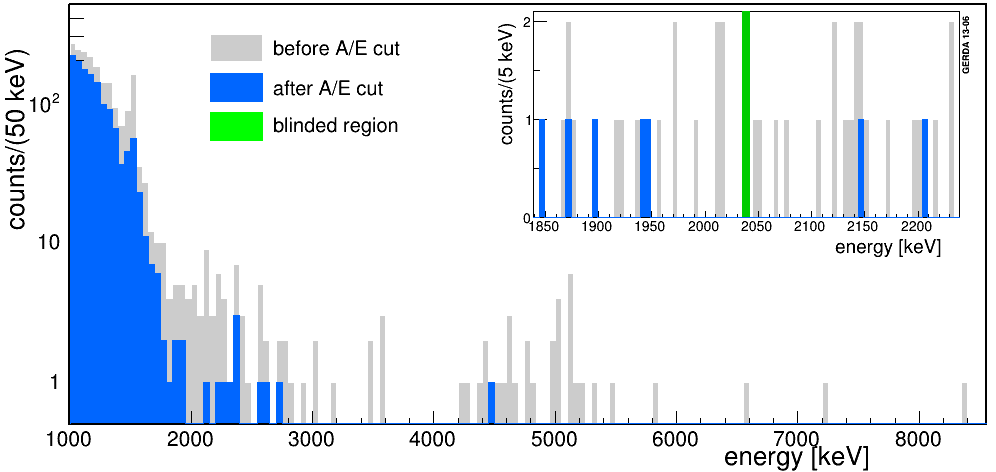}
\caption{Energy spectrum for the \textit{BEGe} data set before (grey) and after
(blue) the PSD cut. A zoom in the $\pm$100 keV window around
Q$_{\beta\beta}$ is shown in the inset. From
Ref. \protect\refcite{gerdapsd}, with kind permission of The European Physical Journal (EPJ).}
\label{fig:psdbegecut}
\end{center}
\end{figure}

\begin{figure}[t!]
\centering
\begin{minipage}[t]{0.45\linewidth}
\includegraphics[width=\textwidth]{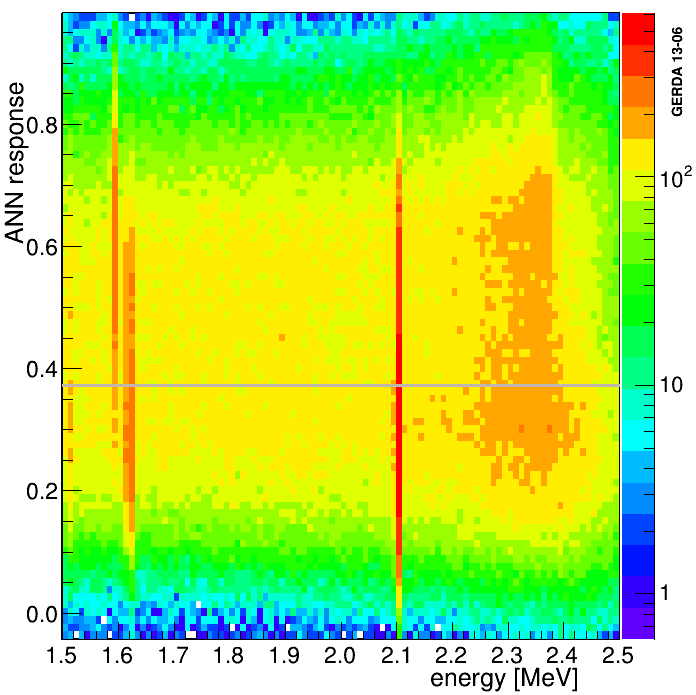}
\caption{Response of the TMlpANN analysis versus energy for events
  from $228$Th calibrations from RG1 detector. The line at $\sim$0.38
  corresponds to the 90\% DEP survival fraction. From
  Ref. \protect\refcite{gerdapsd}, with kind permission of The European Physical Journal (EPJ).}
\label{fig:nnqualifiers}
\end{minipage}
\quad
\begin{minipage}[t]{0.45\linewidth}
\includegraphics[width=\textwidth]{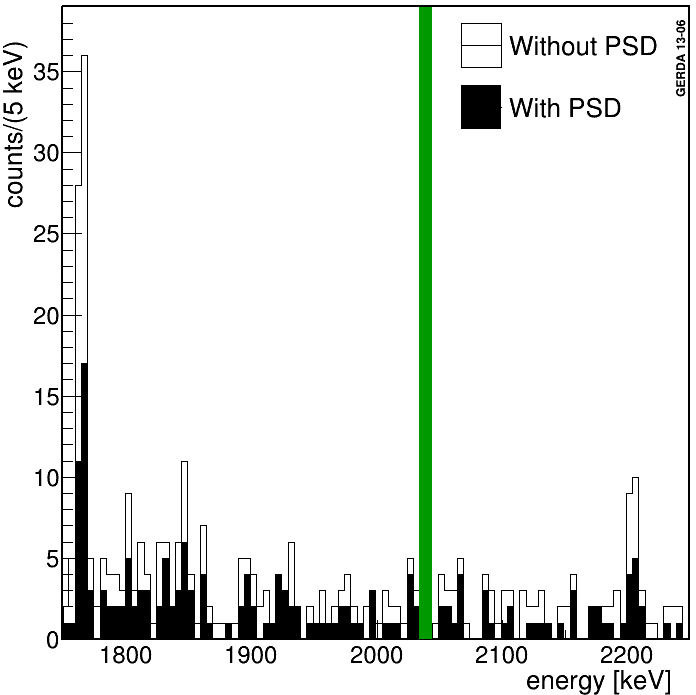}
\caption{Energy
  spectrum from semi-coaxial detectors before (open) and after
  (filled) the PSD selection with TMlpANN. From
  Ref. \protect\refcite{gerdapsd}, with kind permission of The European Physical Journal (EPJ).}
\label{fig:nnspectrum}
\end{minipage}
\end{figure}

In the GERDA detectors, 0$\nu\beta\beta$ events have a peculiar pulse shape which
can be discriminated from background  events. Indeed, the
two electrons from 0$\nu\beta\beta$ decay deposit their energy by
ionization at one location 
in the detectors and those events are called
Single Site Events (SSE). Conversely, the background is mostly due to $\gamma$
induced events and their energy is deposited at multiple locations in the detectors, via multiple Compton
scatterings; the $\gamma$ can, indeed, travel several centimeters. Therefore, such events are called Multi Site Events (MSE). 
The discrimination of 0$\nu\beta\beta$ events, based on the shape
of the recorded pulses, is called Pulse Shape Discrimination (PSD).
In GERDA Phase I two different methods for PSD are used, according
to the different characteristics of the pulses and electricfield
distributions of semi-coaxial and BEGe detectors\cite{gerdapsd}. 
\begin{figure}[t!]
\begin{center}
\includegraphics[width=.78\textwidth]{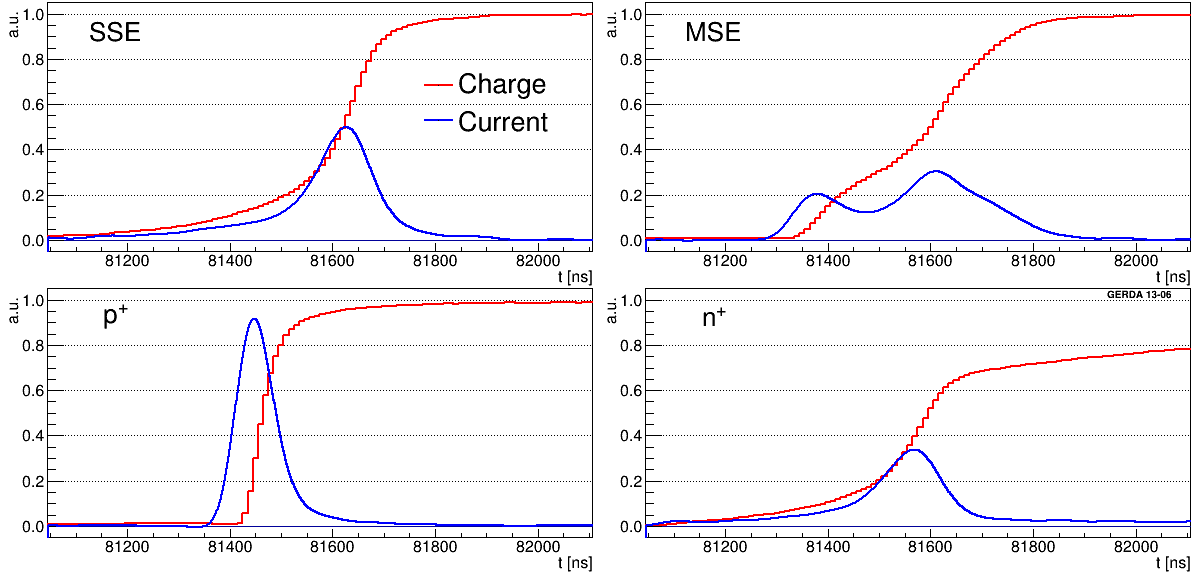}
\caption{Pulse traces from the BEGe detectors. The maximum of the
  charge pulse amplitude is set to 1 and current pulses have equal
  integrals. From Ref. \protect\refcite{gerdapsd}, with kind permission of The European Physical Journal (EPJ).}
\label{fig:bgepulses}
\end{center}
\end{figure}
Concerning BEGes, the ratio between the maximum  {\it A}  of the current
pulse (obtained by numerical differentiation of the charge pulse)
and the energy  \textit{E} of the event (corresponding
to the maximum of the charge pulse) is the discrimination parameter for
Single Site Events with respect to Multi Site Events. This is due to
the fact that in BEGes only holes contribute to the signal and to 
the specific electric field profile; thus holes migrate towards the p$^+$ electrode with very similar paths,
independently from where the energy deposition occurred. 
As a
consequence, for a localized deposition, the maximum of the current pulse and
the energy are proportional. 
In Fig.~\ref{fig:bgepulses}, different examples of
pulse traces and the derived current pulses are shown. 
SSE data (top left) are expected to have a nearly Gaussian
distribution of {\it A/E}, with a width determined by the noise of the
readout electronics. 
The mean of the {\it A/E} values is set to one for the distribution of SSE data. 
In MSE events, current pulses with different
drift times are clearly seen, showing that time-separated pulses are present; in this case
the value of {\it A/E} is below 1. 
In general, for surface events near the p$^+$
electrode, {\it A/E}  is larger than 1 because both electrons and holes contribute; while for {\it n$^+$} surface
events  {\it A/E} is below unity, since only holes contribute and the
current reaches its maximum at the end of the pulse.
The performance of the {\it A/E} based method has been tested with
calibration data. Indeed, in the $^{228}$Th spectrum, 
the double escape peak (DEP) at 1592.5 keV of the 2614.5 keV line from the $^{208}$Tl decay
can be used as a proxy for SSE. The single escape peak (SEP) at 2103.5
keV or full energy peaks (FEP) (like e.g. at 1620.7 keV) represent MSE
data. 
Concerning BEGes, the ratio between the maximum of the current
pulse (obtained by numerical differentiation of the charge pulse) {\it A} 
and the energy of the event \textit{E} (corresponding
to the maximum of the charge pulse) is a discrimination parameter for
Single Site Events with respect to Multi Site Events. This is due to
the fact that in BEGes only holes contribute to the signal and to 
the specific field profile, which causes holes to
migrate towards the p$^+$ electrode with very similar trajectories,
independently from where the energy deposition occurred. As a
consequence, for a localized
deposition, the maximum of the current pulse and
the energy are proportional. 
\begin{figure}[!t]
\begin{center}
\includegraphics[width=.65\textwidth]{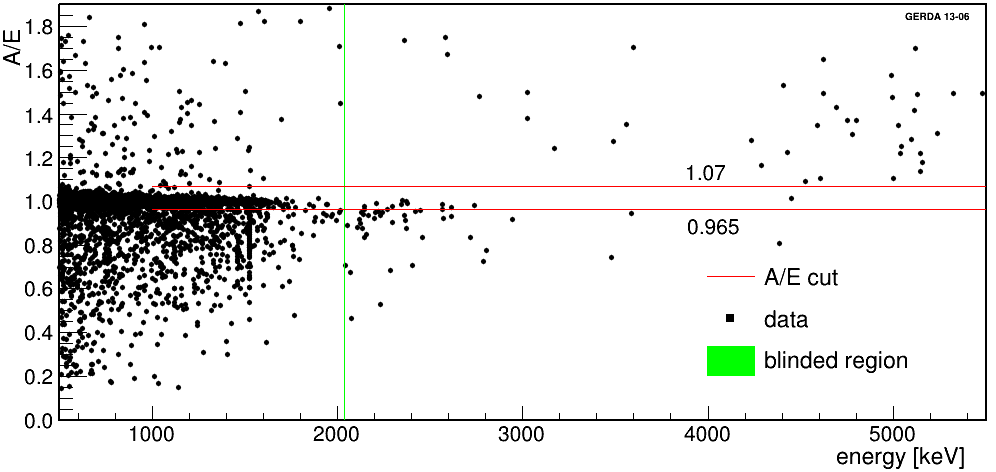}
\caption{{\it A/E} versus energy for the \textit{BEGe} data set. The green band
indicates the blinded region. The red lines indicate the acceptance
region for SSE. From Ref. \protect\refcite{gerdapsd}, with kind permission of The European Physical Journal (EPJ).}
\label{fig:aoveredata}
\end{center}
\end{figure}
In Fig.~\ref{fig:bgepulses}, different examples of
pulse traces and the derived current pulses are shown. 
SSE data (top left) are expected to have a nearly Gaussian
distribution of {\it A/E}, with a width determined by the noise of the
readout electronics. 
{\it A/E} values are rescaled to match the value
of unity at the mean value of the {\it A/E} distribution for SSE data. 
MSE events show current pulses with different
drift times so that time-separated pulses are present; in this case
the value of {\it A/E} is below 1. For surface events near the p$^+$
electrode, 
{\it A/E} is, in general, larger than 1
because, in this case, both electrons and holes contribute, while for {\it n$^+$} surface
events it is below unity, since only holes contribute and the current
peaks at the end of the pulse.
The performance of the {\it A/E} method has been tested with
calibration data. Indeed, in the $^{228}$Th spectrum, 
the double escape peak (DEP) at 1592.5 keV of the 2614.5 keV line from the $^{208}$Tl decay
can be used as a proxy for SSE. The single escape peak (SEP) at 2103.5
keV or full energy peaks (FEP, like e.g. at 1620.7 keV) represent MSE
data. 
No ballistic deficit is observed in the pulses of
SSE events in MC simulations and also when comparing the
reconstructed energy to the risetime of the pulse.
The absence of such
effect is also confirmed by the presence of the DEP line of $^{208}$Tl
at the expected energy in the calibration spectrum (1592$\pm$0.3 keV).
A cross check of the signal detection efficiency is made using
2$\nu\beta\beta$ events, since they are genuine SSE data and  homogeneously
distributed in the detectors,  while DEP events are not.
\\Fig.~\ref{fig:aoveredata} shows {\it A/E} versus energy for the \textit{BEGe}
data set, together with the acceptance region obtained from
data. Accepted events have {\it A/E} between 0.965 (low A/E cut) and
1.07 (high A/E cut). The lower value was determined to have less
than 1\% efficiency losses of the A/E Gaussian function, for energies above 1 MeV. 
Events below the low A/E cut are identified as MSE
and n$^+$ events, while events above the high A/E cut are discriminated
as p$^+$ electrode events.
In Fig.~\ref{fig:psdbegecut} the energy spectrum of BEGe data  is shown before
and after the PSD cut. With a total exposure of 2.4
kg$\cdot$yr for the partially unblinded \textit{BEGe} data set, seven out of 40 events survive the
cut in the 400 keV region around Q$_{\beta\beta}$ (excluding the 8 keV
blinded window) and the BI is reduced from 0.042$\pm$0.007 to
0.007$^{+0.004}_{-0.002}$ cts/(keV$\cdot$kg$\cdot$yr). 
The acceptance
efficiency for signal-like events (i.e. the survival fraction of
0$\nu\beta\beta$ events) is 0.92$\pm$0.02 and it is derived from the
survival fraction of DEP events and from Monte Carlo simulations of
the 0$\nu\beta\beta$ signal.
The fraction of
background events, rejected at Q$_{\beta\beta}$, is about 80\%. 
The method shows, therefore, a very good performance, with
  both high background reduction and high acceptance efficiency.
\\For semi-coaxial detectors, the A/E parameter does not represent a
useful variable for pulse shape discrimination. Different shapes of
the current pulses are, in fact, observed in the bulk volume,
moving from the outer n$^+$ surface to the p$^+$ surface, 
due to the contribution of both electrons and holes to the pulse.
Instead, a method based on the artificial neural network for the
rising part of the charge pulses was applied. It is based on
the TMlpANN\cite{tmlpann} algorithm (implemented in
the TMVA\cite{tmva} toolkit of ROOT) based on multilayer
perceptrons and on the so called ``supervised learning'' algorithm. 
Two hidden layers with 50 and 51 neurons were used. 
The times $t_1 \, t_2 \, ,  \cdots, t_n$, at which the 1,3,5,...,99\%
of the full height of the pulse is reached, were used as input
parameters. Being the sampling frequency in GERDA 100 MHz, two
consecutive time bins were interpolated. Calibration data were used for
the network's training; data at the DEP peak were considered as SSE and those at the
full line of $^{212}$Bi (1621 keV) were considered as MSE. The method
has been tuned to give 90\% survival fraction for DEP 
events from the gamma line of $^{208}$Tl decays at 2.6 MeV.
The output of the neural network is a qualifier, i.e. a number between $\sim$0 (background like events) and $\sim$1 (signal
like events). 
Fig.~\ref{fig:nnqualifiers} shows a scatter plot of
this variable as a function of the energy. The qualifier threshold for
90\% survival probability of DEP events was determined for each detector and
each time period considered. The possible deviations from 0.90, due to
an energy dependence or to a volume effect, associated to different
contributions from DEP and 0$\nu\beta\beta$ events, were combined
quadratically. The final value for the 0$\nu\beta\beta$ efficiency is
0.90$^{+0.05}_{-0.09}$. 
The method rejects about 45\%
of the events in the 230 keV window around Q$_{\beta\beta}$. The energy spectrum
before and after the PSD cut is shown in Fig.~\ref{fig:nnspectrum}.
\\Two additional methods were used to 
cross check the results from neural network analysis. 
The first is based on a likelihood method 
and the second relies on the correlation between the A/E
parameter and the pulse asymmetry. 
 In the 230 keV window around Q$_{\beta\beta}$, about 90\% of the events rejected by the 
neural network method were also rejected by the two other analyses.  

\section{Limit on the half-life of 0$\nu\beta\beta$ decay in
  $^{76}$Ge}
\label{sec:0nbb}
The total collected exposure of GERDA Phase I data (21.6 kg$\cdot$yr) was considered to derive
a limit on the half-life of 0$\nu\beta\beta$ decay in $^{76}$Ge\cite{0nbbpaper}. 
Data were collected from November 2011 to May 2013, corresponding to 492.3
live days. A fraction of 5\% of the data was discarded due to
temperature instabilities. Data were processed offline according to the procedure
described in Ref. ~\refcite{ref20paper0nbb}, implemented in the
software tool GELATIO\cite{gelatio}. The reconstruction of the deposited energy is
made by a digital filter with semi-Gaussian shaping. The energy scale
of the individual detectors is determined by weekly calibrations with
$^{228}$Th sources. The exposure-weighted average energy resolution (FWHM),
extrapolated at Q$_{\beta\beta}$, is (4.8$\pm$0.2) keV for semi-coaxial
detectors and (3.2$\pm$0.2) keV for BEGes. 
The energy resolution of the detectors is slightly degraded
 with respect to the value determined by the HdM experiment. Indeed, the detectors are
placed directly in LAr and not in a standard vacuum cryostat and also
there is a relatively long distance (about 30 cm) between the diode and the front-end
electronics. 
Of course a better resolution would result in an even better sensitivity for the half-life of the
0$\nu\beta\beta$ decay. Recent studies to improve the resolution of
GERDA detectors are very promising~\cite{taupposter}. 
\\A blind analysis was performed to avoid biases in the event selection
criteria. The subdivision of data into subsets and the blinding procedure
have already been described in Sect.~\ref{sec:gerdabackground}.
Different analysis cuts were applied to discard possible background
signals: (i) only events with energy deposition in a single detector
are accepted (anti-coincidence cut). This cut reduces background
around Q$_{\beta\beta}$ of about 15\%. (ii) events from detectors
 in coincidence within 8 $\mu$s with a signal from muon veto
are rejected. An additional 7\% reduction of the background is
obtained. (iii) events preceded or followed by an other event within 1
ms are rejected. This cut rejects events from the $^{214}$Bi-$^{210}$Po
cascade in the $^{222}$Rn decay chain. The background reduction by
this cut is less than 1\%. 
\\In addition to the previous cuts, pulse shape discrimination (described
in Sect.~\ref{sec:psd}) was applied. 
\begin{figure}[t!]
\begin{center}
\includegraphics[width=.65\textwidth]{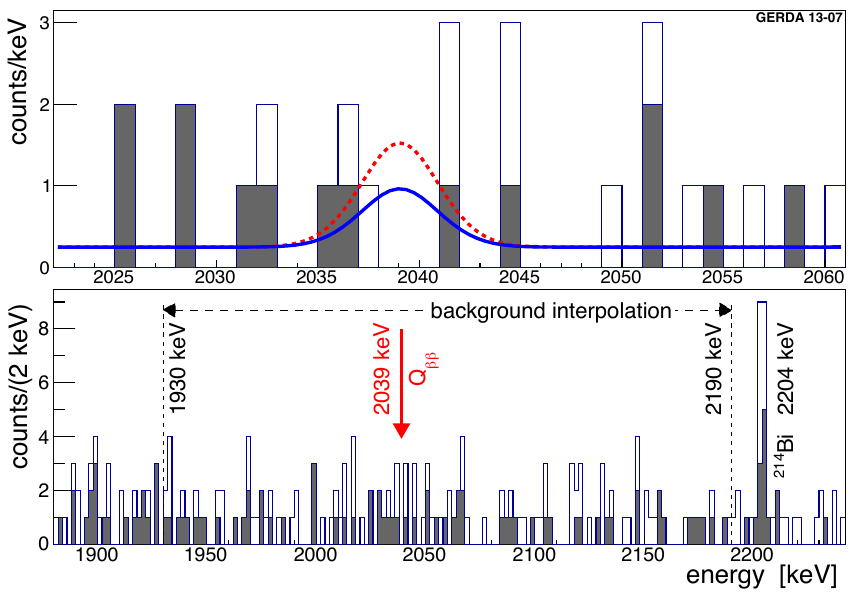}
\caption{Energy spectrum from all $^{enr}$Ge detectors with (filled)
  and without (open) the PSD selection. In the upper panel 
the expectation based on the central value of the half-life
  predicted by Ref. \protect\refcite{hdmclaim} is also shown (red), together with the 90\%
  C.L. limit predicted by GERDA Phase I (blue). In the lower panel the energy
  window used for the backgrund interpolation is indicated. Plot from
  Ref.~\protect\refcite{0nbbpaper}.}
\label{fig:0nbb}
\end{center}
\end{figure}
The total energy spectrum, 
before and after the PSD selection, is shown in figure
\ref{fig:0nbb}. 
The background is flat in the
Q$_{\beta\beta}$$\pm$5 keV range and seven events are observed while
5.1$\pm$0.5 are expected from background counts. After the PSD cut, 
three of the six events from the semi-coaxial detectors and the one
from the BEGe detector were classified
as background. 
No event remained in the energy window Q$_{\beta\beta}$$\pm
\sigma_E$ and, therefore, no excess of events was present.
\\The half-life on the 0$\nu\beta\beta$ decay is calculated
according to Eq.~\ref{eq:t12n0nu}.
For GERDA, the efficiency factor $\epsilon$ contains the following terms:
\begin{equation}
\epsilon = f_{76} \cdot f_{av} \cdot \epsilon_{fep} \cdot \epsilon_{psd}
\label{eq:efficiencies}
\end{equation}
where $f_{76}$ is the fraction of $^{76}$Ge atoms in Ge, $f_{av}$ is the active volume
fraction, $\epsilon_{fep}$ is the
probability for a 0$\nu\beta\beta$ decay to release its entire energy
into the active volume and $\epsilon_{psd}$ is the efficiency of the PSD
analysis. 
\\The analysis to derive the signal strength was performed according to
a profile likelihood fit on the three GERDA data sets. The fitted function
contains three constant terms for the background from the three data
sets and a Gaussian peak, centered at Q$_{\beta\beta}$ 
and with standard deviation equal to the energy resolution (FWHM). The four
corresponding parameters of the function were the three terms for the
background and 1/T$^{0\nu}_{1/2}$, the latter being proportional to the peak 
counts (see Eq.~\ref{eq:t12n0nu}) and common to the three subsets.
The best fit value obtained is N$^{0\nu}$=0 pointing out that no excess above
background is found. The limit on the half-life is 
\begin{equation}
T^{0\nu}_{1/2} > 2.1 \cdot 10^{25} \mbox{ yr } (90\% \mbox{ C.L.}).
\end{equation}
\begin{figure}[t!]
\begin{center}
\includegraphics[width=.55\textwidth]{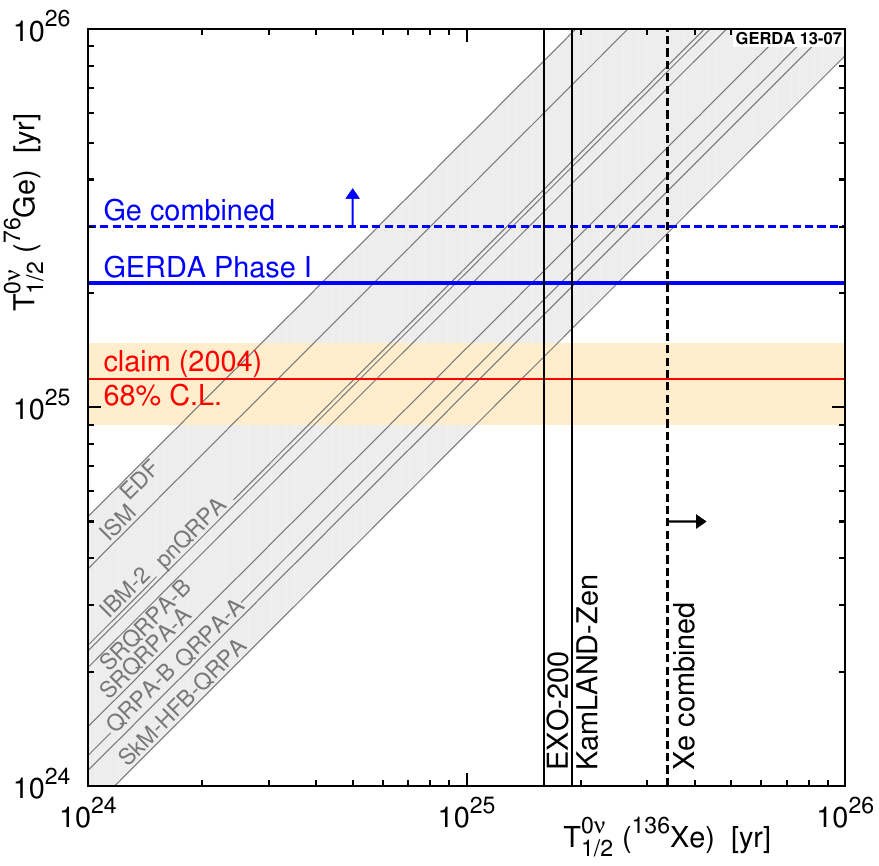}
\caption{90\% C.L. limits on T$^{0\nu}_{1/2}$ for $^{76}$Ge and
  $^{136}$Xe, compared with the signal claim of Ref. \protect\refcite{hdmclaim}. The
  shaded gray bands indicate the predictions for the correlation of
  half-lives in $^{76}$Ge and $^{136}$Xe, according to the different NME calculations
considered. Plot from Ref. \protect\refcite{0nbbpaper}.}
\label{fig:gexe}
\end{center}
\end{figure}
The systematical uncertainties due to detector
parameters, selection efficiency, energy resolution and energy scale,
were folded into the half-life estimation; they weaken the limit by about 1.5\%.
The corresponding limit on the number of signal events is N$^{0\nu} <$
3.5 counts. The median sensitivity for the 90\% C.L. limit, given the
background levels and the efficiencies, is T$^{0\nu}_{1/2} >$
2.4 $\cdot$ 10$^{25}$ yr. A Bayesian analysis\cite{likelihood}
was also performed (using the BAT toolkit\cite{bat}) 
with the same fit and a flat prior distribution for
1/T$^{0\nu}_{1/2}$ between 0 and 10$^{-24}$ yr$^{-1}$. The
corresponding result for the limit on the half-life is $T^{0\nu}_{1/2}
>$ 1.9 $\cdot$ 10$^{25}$ yr, with a median sensitivity of $T^{0\nu}_{1/2}
>$ 2.0 $\cdot$ 10$^{25}$ yr.
\\ The GERDA result does not support the previous claim of
0$\nu\beta\beta$ decay observation in $^{76}$Ge\cite{hdmclaim}. Rescaling the
number of counts corresponding to the half-life from Ref.~\refcite{hdmclaim},
GERDA should observe (5.9$\pm$1.4) 0$\nu\beta\beta$ decays at Q$_{\beta\beta} \pm$ 2$\sigma_E$
with (2.0$\pm$0.3) background events, while only 3 events were actually observed in
this energy window. The Bayes factor, i.e. the ratio between the
posterior probability of the model H$_1$ (assuming the value of
T$^{0\nu}_{1/2}$ from Ref. ~\refcite{hdmclaim}) and the posterior probability of
the model H$_0$ (assuming only background), is P(H$_1$)/P(H$_0$) = 0.024.
\\The limit found by GERDA is consistent with the limits found from the HdM
and IGEX experiments. A combined profile likelihood of the three
results gives N$^{0\nu}$ = 0 as best
fit and:
\begin{equation}
T^{0\nu}_{1/2} > 3.0 \cdot 10^{25} \mbox{ yr } (90\% \mbox{ C.L.}).
\end{equation}
A Bayesian analysis gives the same limit and a Bayes factor of
P(H$_1$)/P(H$_0$) = 2$\cdot$10$^{-4}$. 
\\Results from $^{76}$Ge experiments can be compared to the recent
limits from KamLAND-Zen\cite{kamland} and EXO-200\cite{exo} on
$^{136}$Xe half-life, assuming that the 0$\nu\beta\beta$ process is due to
the exchange of a light Majorana neutrino.
In this case the ratio of the 0$\nu\beta\beta$ half-lives 
is proportional to the square of the ratio between the nuclear matrix
elements $M_{0\nu}$($^{76}$Ge)/$M_{0\nu}$($^{136}$Xe). 
In Fig.~\ref{fig:gexe} the limits set on T$^{0\nu}_{1/2}$ for
$^{76}$Ge and $^{136}$Xe by the different experiments are shown,
together with different NME calculations and the limit found from the combination
of GERDA, KamLAND-Zen and EXO-200 results.
Considering the most conservative value
for the NME, the Bayes Factor obtained by this combination 
is 0.0022. The claim is again strongly disfavoured. It is worth to note that
other theoretical calculations could lead to smaller NME ratios and, consequently,
to weaker conclusions. 
\\Considering the most recent value for the $^{76}$Ge
phase-space factor\cite{wavefunctions} 
and the NME calculations reported in
Refs. from~\refcite{nmecalc1} to~\refcite{nmecalc7} 
(scaling the different $g_A$ and $R_A$
parameters according to Ref. \cite{peter}),
the derived upper limits on the effective electron neutrino mass 
range between 0.2 and 0.4 eV.

\section{Conclusions}
The Gerda experiment has completed the Phase I with a total collected
exposure of 21.6 kg$\cdot$yr. No events from 0$\nu\beta\beta$ decay
have been observed and a lower limit on the half-life on
the 0$\nu\beta\beta$ decay for $^{76}$Ge has been estimated to be T$^{0\nu}_{1/2} >$
2.1 $\cdot$ 10$^{25}$ yr at 90\% C.L.
The previous claim for a 0$\nu\beta\beta$ signal\cite{hdmclaim}
(T$^{0\nu}_{1/2}$ = 1.19 $\cdot$ 10$^{25}$ yr) is 
strongly disfavoured by the GERDA result. The GERDA result was not to
compared to the value T$^{0\nu}_{1/2}$ = 2.23 $\cdot$ 10$^{25}$ yr obtained from the re-analysis of HdM data 
because of some inconsistencies in the analysis already pointed out in Ref.~\refcite{bernhard}.
In the future Phase II of GERDA, the expected sensitivity on the half-life for
0$\nu\beta\beta$ decay will be about 10 times higher than Phase I,
T$^{0\nu}_{1/2} >$ 10$^{26}$ yr; as a result, lower values of the effective
Majorana neutrino mass will be explored.

\end{document}